\documentclass{emulateapj}
\usepackage{txfonts}

\slugcomment{ApJL, in press}

\shorttitle{Radio--X-ray Flare in Cygnus~X-1}
\shortauthors{Wilms et al.}

\begin{document}

\title{Correlated radio--X-ray variability of Galactic Black Holes: A
  radio--X-ray flare in Cygnus~X-1}

\author{
J\"orn Wilms\altaffilmark{1},
Katja Pottschmidt\altaffilmark{2},
Guy G. Pooley\altaffilmark{3},
Sera Markoff\altaffilmark{4},
Michael A. Nowak\altaffilmark{5},
Ingo Kreykenbohm\altaffilmark{6,7},
Richard E. Rothschild\altaffilmark{2}
}

\altaffiltext{1}{Dr.\ Karl Remeis-Observatory, University of
  Erlangen-Nuremberg, Sternwartstr.~7, 96049 Bamberg, Germany, joern.wilms@sternwarte.uni-erlangen.de}
\altaffiltext{2}{Center for Astrophysics and Space Sciences,
  University of California, San Diego, La Jolla, CA 92093-0424, USA,
  kpottschmidt@ucsd.edu, rrothschild@ucsd.edu}
\altaffiltext{3}{Mullard Radio Astronomy Observatory, Cavendish
  Laboratory, Madingley Road, Cambridge CB3 0HE, UK, guy@mrao.cam.ac.uk}
\altaffiltext{4}{Astronomical Institute ``Anton Pannekoek'',
  University of Amsterdam, Kruislaan 403, Amsterdam, 1098 SJ, The
  Netherlands, sera@astro.uva.nl} 
\altaffiltext{5}{MIT Kavli Institute for Astrophysics and Space
  Research and \textsl{Chandra} X-ray Center, NE80-6077, 77
  Massachusetts Ave., Cambridge, MA 02139, USA, mnowak@alum.mit.edu}
\altaffiltext{6}{Institut f\"ur Astronomie und Astrophysik --
  Astronomie, Sand 1, 72076 T\"ubingen, Germany,
  kreyken@astro.uni-tuebingen.de}
\altaffiltext{7}{\textsl{INTEGRAL} Science Data Centre, Chemin
  d'\'Ecogia 16, 1290 Versoix, Switzerland}

\email{joern.wilms@sternwarte.uni-erlangen.de}

\begin{abstract}
  We report on the first detection of a quasi-simultaneous
  radio--X-ray flare of Cygnus~X-1. The detection was made on 2005
  April 16 with pointed observations by the Rossi X-ray Timing
  Explorer and the Ryle telescope, during a phase where the black hole
  candidate was close to a transition from the its soft into its hard
  state. The radio flare lagged the X-rays by $\sim$7\,minutes,
  peaking at 3:20\,hours barycentric time (TDB 2453476.63864). We
  discuss this lag in the context of models explaining such flaring
  events as the ejection of electron bubbles emitting synchrotron
  radiation.
\end{abstract}


\section{Introduction}
With the increased availability of simultaneous radio and observations
in the last decade, there is now a large amount of evidence available
pointing towards a very close interaction between the accretion disk
and the jet in black hole X-ray binaries and active galactic nuclei
(AGN). Most convincingly, this disk-jet interaction has been shown for
microquasars, i.e., black hole binaries with strongly relativistic
jets such as \object{GRS~1915+105} or \object{GRO J1655$-$40}. In
these systems, the correlated flaring in the X-rays,
optical/infra-red, and radio seen at certain times is generally
interpreted as the evidence for (ballistic) ejection events of
synchrotron radiation-emitting electron bubbles \citep[][ and
therein]{rothstein:05a,fender:04a,kleinwolt:02a,eikenberry:98a}. In
this model, the X-ray flare represents the ejection of the synchrotron
radiation emitting bubble, which then adiabatically expands within the
jet flow and cools down, resulting in the peak of the emission
shifting downwards in frequency with time
\citep{vanderlaan:66a,hjellming:88a}.  Simultaneous broadband
observations of such events, which show minute-long delays between the
different wave bands, are consistent with this picture
\citep{mirabel:94a,mirabel:98a,pooley:97a,eikenberry:98a}. The model
has also been confirmed by proper motion measurements in the radio,
which reveal intrinsic jet speeds of $\gtrsim$$0.57c$ for GRS~1915+105
\citep{millerjones:05a}.  Comparable behavior was also detected in
\object{3C\,120}, suggesting that similar ejections also occur in
active galactic nuclei, on correspondingly longer timescales
\citep{marscher:02a}.

For black hole binaries with weakly relativistic jets, the evidence
for jet-disk-interaction is less direct.  This evidence includes the
correlation between X-ray states and radio emission in black hole
transients \citep[e.g., in
\object{GX~339$-$4}][]{corbel:03a,belloni:05a} and the success of
modeling the radio to X-ray broad band spectrum of black hole
candidates with outflow-dominated models \citep[][ and
therein]{markoff:05a,markoff:04a}. Furthermore, at least for
\object{Cygnus~X-1}, there is also evidence for the presence of an
energetically significant, strong outflow
\citep{stirling:98a,stirling:01a,gallo:05a,millerjones:06a}. A
relativistic jet with $v\gtrsim 0.3c$ has been associated with radio
flares in this system \citep{fender:06a}.

Apart from GRS~1915+105, however, none of these observations shows
direct evidence for a causal connection between the X-rays and jet on
time scales of minutes.  Prompted by this lack of quasi-simultaneous
short-term radio--X-ray correlations, in 1998 we initiated a long term
monitoring campaign of Cyg~X-1 with the Rossi X-ray Timing Explorer
(\textsl{RXTE}) and the Ryle telescope. Biweekly 3--10\,ksec long
simultaneous observations started in 1999.  Previous searches for
flares in campaign data taken between 1999 and mid-2003 did not reveal
evidence for coherent short term activity in both bands, although a
significant correlation on time scales of weeks was found, especially
above $\sim$10\,keV \citep{gleissner:04a,wilms:05a}.  In this
\emph{Letter}, we report on the observation made on 2005 April 16, in
which the first clear quasi-simultaneous radio--X-ray flare was
detected in Cyg~X-1.  The remainder of this \emph{Letter} is
structured as follows.  In \S\ref{sec:obs} we describe the
observations, followed by the analysis of the flare in
\S\ref{sec:flare}. We discuss the physics of the flare in the context
of emission models for the radio and X-ray emission in
\S\ref{sec:discuss}.

\section{Observations and Data Reduction}\label{sec:obs}
We use data from both instruments on-board the \textsl{RXTE}, the low
energy Proportional Counter Array \citep[PCA;][]{jahoda:05a} and the
High Energy X-ray Timing Experiment \citep[HEXTE;][]{rothschild:98a}.
The data analysis was performed using the standard \textsl{RXTE} data
analysis software, HEASOFT 6.1.2.  Spectral fitting was performed with
XSPEC 11.3.2aa \citep{arnaud:96a}.

A crucial part of the observation happened during the early phase of
the \textsl{RXTE} observation, shortly after the source rose above the
Earth's horizon. Due to auroral emission in the far ultra-violet and
soft X-rays and due to cosmic ray reprocessing in the hard X-rays, the
Earth's atmosphere is not completely X-ray dark. The typical
2.5--20\,keV X-ray flux at the typical magnetic latitude of the
\textsl{RXTE} orbit is too low, however, to influence our measurements
\citep{petrinec:00a,sazonov:07a}. We therefore use all data taken
whilst the source was $\ge$$1^\circ$ above the Earth's horizon and had
a source offset of $\le$$0\fdg{}01$.  We use PCA data from the
standard1 mode, which gives the full 2.5--128\,keV PCA count rate with
a time resolution of 0.125\,s and no energy information, and from the
standard2f mode, a binned data mode with a 128 channel energy
resolution and a time resolution of 16\,s.  X-ray light curves were
extracted with the intrinsic time resolution of each mode and then
barycentered and rebinned.

The Ryle telescope data were taken at 15\,GHz with a time resolution
of $\sim$8\,s. The typical $1\sigma$ uncertainty of the radio
measurements is 9\,mJy. The observations are interrupted every
$\sim$1600\,s for phase calibration observations of J2007+4029. The
amplitude calibration of the Ryle data corresponds to the flux scale
of \citet{baars:77a} and is performed using nearby observations of
3C48 and 3C286. See \citet{pooley:97a} for further information on the
Ryle telescope.

\begin{figure}
\plotone{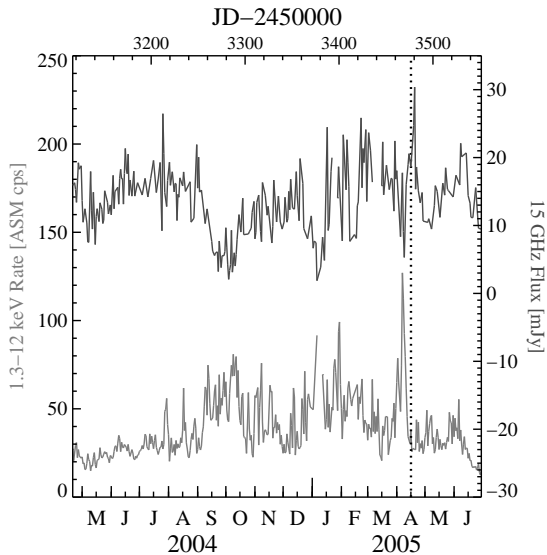}
\caption{\textsl{RXTE}-ASM (lower curve, left $y$-axis) and Ryle telescope
  15\,GHz (upper curve, right $y$-axis) lightcurves of Cyg~X-1 from
  2004 March until 2005 June. The time of the simultaneous observation
  is indicated by the dashed vertical line. Gaps are shown if their
  duration is $\ge$4\,d.}\label{fig:longterm}
\end{figure}

\section{A Quasi-Simultaneous Radio--X-ray Flare}\label{sec:flare}

\begin{figure}
\plotone{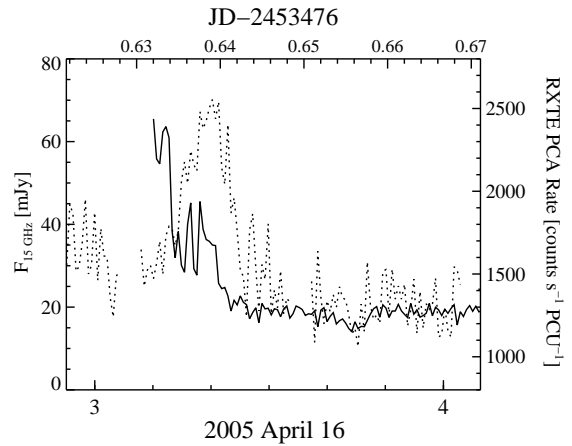}
\caption{Barycentric 15\,GHz flux density (left axis, dotted line) and
  2--128\,keV count rate lightcurves (right axis, solid line) of
  Cyg~X-1 for the flare event of 2005 April 16, both binned to a
  resolution of 32\,s. The bottom axis gives the barycentric time in
  hours, with tick-marks separated by 10\,minutes.}\label{fig:lc}
\end{figure}

As shown in Fig.~\ref{fig:longterm}, 2005 April marks the possible end
of a longer X-ray flaring episode of Cyg~X-1 that started in early
2004 \citep{wilms:05a}.  While clearly defined radio flares are not
uncommon in Cyg~X-1 \citep[e.g.,][]{hjellming:73a}, increased radio
emission and radio flaring are generally seen when the source is in
the intermediate state between the hard and the soft states, while the
radio is weak once the X-ray source approaches the soft state
\citep[][ and therein]{wilms:05a}.  At the time of our pointed
observations, the soft X-ray flux had just come down from a large
flare.  Shortly after the observation, the 1\,d averaged 15\,GHz flux
peaked, reaching a maximum of $\sim$30\,mJy, close to the brightest
radio flux of Cyg~X-1 during 2004/2005.

Figure~\ref{fig:lc} shows the 15\,GHz radio flux and the \textsl{RXTE}
PCA count rate lightcurve measured on 2005 April 16. Close to the
start of the observation, a radio flare is readily apparent. The total
duration of the flare is $\sim$15\,minutes. During this interval the
15\,GHz flux increased by a factor of $\sim$3 to a peak radio flux of
70\,mJy. This radio flux is among the highest seen during the Ryle
monitoring\footnote{The most exceptional radio flare was that of 2004
  February 20, which reached a peak flux of 140\,mJy at 15\,GHz, the
  largest flux ever seen for this source with the Ryle telescope
  \citep{fender:06a}.}. Previous radio flares, however, did not occur
during pointed \textsl{RXTE} observations
\citep{gleissner:04a,fender:06a}, and the source monitoring provided
by the \textsl{RXTE}-ASM is too coarse to pick up such short lived
X-ray events.

\begin{figure}
\plotone{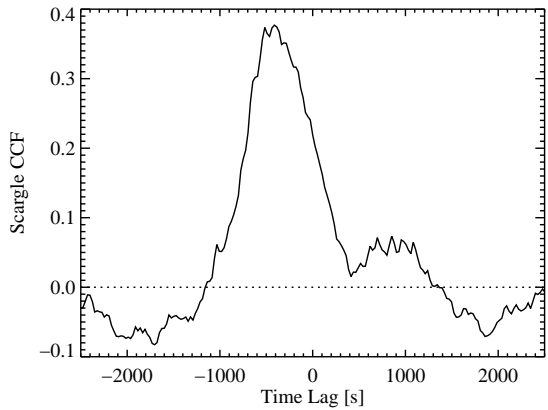}
\caption{Cross-correlation function \citep{scargle:89a} for the radio
  data with respect to the \textsl{RXTE}-PCA data. A negative time lag
  indicates the radio lagging the X-rays.}\label{fig:corr}
\end{figure}

\textsl{RXTE} started observing Cyg~X-1 about 10 minutes before the
peak radio flux. The X-ray lightcurve shown in Fig.~\ref{fig:lc} shows
a similar shape to the radio one, although with more substructure.
The earlier maximum of the X-ray flare did not allow \textsl{RXTE} to
catch the start of the X-ray flare, or determine whether the maximum
X-ray flux seen is indeed the peak of the X-ray flare.  A cross
correlation (CCF) analysis using the algorithm of \citet{scargle:89a}
reveals a $413\pm 165$\,s time lag of the radio with respect to the
X-rays, where the $1\sigma$ uncertainty was determined using a
standard bootstrapping method with 1000 realizations.  Other
approaches to calculate the CCF for non-uniformly sampled data
\citep{alexander:97a,edelson:88a} give essentially the same result.
With a maximum \citet{scargle:89a}-CCF of 0.38, this analysis formally
confirms the general similarity of the X-ray and radio lightcurves
(Fig.~\ref{fig:corr}).  Since the substructure of the X-ray
lightcurve, i.e., the two smaller flares after the main flare, is
clearly different from that in the radio flare and since the start of
the X-ray flare is not covered by our observations, the peak CCF value
is not higher.  For the same reasons, the formal uncertainty of the
lag measurement is rather large.

To characterize the shape of the radio flare, we fit the radio data
(rebinned to a resolution of 8\,s) with the sum of a linear flux trend
and a Gaussian representing the flare,
\begin{equation}\label{eq:flarefunc}
  f(t) = a (t-t')+b + A \exp\left( -\frac{(t-T)^2}{\sigma^2}\right) 
\end{equation}
where $t'$ is a reference time, taken as the center of the time
interval analyzed ($t'={\rm JD}\,2453476.64635$). The barycentric time
of maximum flux occured at $T={\rm JD}\,2453476.63864$ with an
uncertainty of 15\,s (uncertainties are at the $1\sigma$ level for one
interesting parameter). The peak flux of the flare component is
$A=39.2\pm2.5$\,mJy and the width of the component is $\sigma=210\pm
16$\,s.  The flare is superimposed to a continuum with
$b=25.3\pm0.7$\,mJy, decreasing linearly with $a=-13\pm2\,{\rm
  mJy}\,{\rm h}^{-1}$.  The high quality of the fit is indicated by
its low reduced $\chi^2$ ($\chi^2_{\rm red}=1.11$ for 233 degrees of
freedom). The radio flare is therefore symmetric around its maximum.
This symmetry is a marked difference compared to the asymmetric flare
of 2004 February 20 \citep{fender:06a}.

Modeling the \textsl{RXTE}-PCA light curve is complicated by the flare
already being in progress when the measurements started.  Furthermore,
contrary to the radio data, where the scatter in the lightcurve is
mainly due to the measurement uncertainty, the X-ray data are
dominated by strong low-frequency noise onto which the X-ray flare is
superimposed.  Consequently, the empirical model of
Eq.~(\ref{eq:flarefunc}) does not result in a good description of the
X-ray data.

To study the spectral evolution of Cyg~X-1 during the flare, we
perform a spectral analysis of the 2.5--20\,keV PCA standard2f data at
16\,s time resolution using a simple photoabsorbed powerlaw, which
proves sufficient to describe the spectrum at this lower signal to
noise level. For spectra taken during the flare, at PCA count rates
above $1000\,{\rm counts}\,{\rm s}^{-1}\,{\rm PCU}^{-1}$, the mean
power law index $\Gamma=2.10\pm 0.03$. The spectrum hardens outside of
the flare to $\Gamma=1.98\pm 0.03$ (errors given are the standard
deviation of the individual power law fits to the standard2f spectra),
a value typical for the intermediate state of this source \citep{wilms:05a}.

That Cyg~X-1 was in the intermediate state on the day of the flare can
also be confirmed by modeling the 2.5--150\,keV PCA and HEXTE spectrum
of an \textsl{RXTE} observation performed 4\,h after the flare (to
avoid possible ``contamination'' by the flaring activity) with the sum
of a photoabsorbed ($N_{\rm H}=6\times 10^{21}\,{\rm cm}^{-2}$, held
fixed), exponentially cut-off broken power law model. This empirical
model has been shown to give a good characterization of the spectral
shape of Cyg~X-1 \citep{wilms:05a}.  The spectral parameters are a
lower photon index $\Gamma_1=2.01\pm 0.01$, breaking at $E_{\rm
  break}=10.0^{+0.3}_{-0.2}\,{\rm keV}$ into a power law with
$\Gamma_2=1.62^{+0.02}_{-0.03}$. At $E_{\rm cut}=26\pm3\,{\rm keV}$
the exponential cutoff starts with a folding energy of $E_{\rm
  fold}=137\pm 11\,{\rm keV}$ (all uncertainties are at the 90\%
level). In addition, a Fe K$\alpha$ line from neutral iron is present
with an equivalent width of 135\,eV.  The 3--10\,keV source flux is
$6.17\times 10^{-9}\,{\rm ergs}\,{\rm cm}^{-2}\,{\rm s}^{-1}$.  The
parameters of the continuum are again consistent with an intermediate
state and fits well with the empirical picture that radio flaring in
black hole candidates occurs most frequently in this state
\citep{fender:04b,wilms:05a}.

\begin{figure}
\plotone{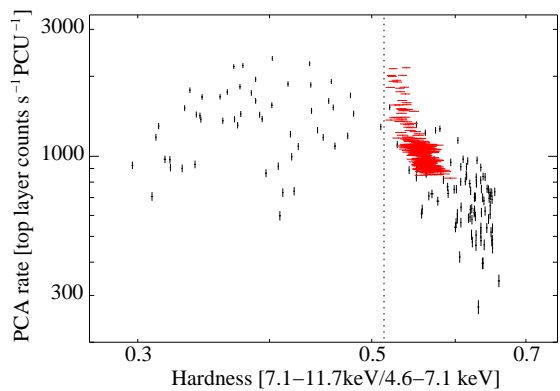}
\caption{1--128\,keV \textsl{RXTE}-PCA count rate versus hardness
  ratio for Cyg X-1 observations from the 2000--2005 \textsl{RXTE}
  monitoring (black) and for data taken during the 2005 April 16 flare
  (red). The dotted line indicates the location at which observations
  with maximum radio flux are found in the long term monitoring. The
  hardness ratio was determined from background subtracted PCA top
  layer data from channels 8--14 and 15--25,
  respectively. \textit{A color version of this figure is available in
  the online edition of ApJL.}}\label{fig:hardness_int}
\end{figure}

To allow the interpretation of the observed softening during the flare
with the general behavior of Cyg~X-1, Fig.~\ref{fig:hardness_int}
shows the X-ray hardness intensity diagram for the PCA top layer
standard2f 16\,s spectra in the context of the pointed \textsl{RXTE}
observations of the monitoring campaign. For black hole transients
this diagram is seen to have an approximate \textsf{q}-shape \citep[][
and therein]{belloni:06a}. As a persistent hard state source, Cyg~X-1
is typically found in the top right corner of the diagram.  Outside of
the flare, the source is situated at a hardness of $\sim$0.55 with a
typical 1--128\,keV PCA count rate of
$\sim$$900\,{\rm counts}\,{\rm s}^{-1}\,{\rm PCU}^{-1}$. During the
flare the source softens and brightens. It leaves the region of the
diagram where Cyg~X-1 is usually found during our monitoring campaign,
by moving to higher count rates, for this hardness, than usually
observed.

Flaring behavior in Cyg~X-1 is usually observed whenever the source is
close to the ``jet line'' in its hardness intensity diagram, while the
radio flux gets quenched once the source moves away from the jet line
to the left of the diagram \citep{gallo:03a,fender:04b}. In spectral
fits based on the \texttt{eqpair}-model of \citet{coppi:99a}, the
radio fluxes of Cyg~X-1 are at their maximum when the compactness
ratio $\ell_{\rm h}/\ell_{\rm s}\sim 3$ \citep[Fig.~16 of][; the
compactness ratio is a measure for the relative importance of the
energy dumped into the Comptonizing plasma and that dissipated in the
accretion disk]{wilms:05a}, corresponding to a soft power law index of
$\Gamma_1\sim 2.1$. From our database of spectra of Cyg~X-1, we find
that these observations have a $(7.1\mbox{--}11.7\,{\rm
  keV})/(4.6\mbox{--}7.1\,{\rm keV})$ hardness ratio of 0.52,
indicated by the dotted line in Fig.~\ref{fig:hardness_int}.
During the 2005 April 16 observation, the source was therefore close
to this line of maximum radio flux and approached it asymptotically
during the flare.

\section{Discussion}\label{sec:discuss}

In this \emph{Letter} we have presented the first evidence for a
direct relationship of the X-ray and radio emission in Cyg~X-1 on
timescales of minutes. The data show the radio to lag the X-rays by
$413\pm 165$\,s and the X-ray spectral shape to approach the X-ray
hardness ratio where the source is typically found at its largest
radio flux in our long term monitoring, hinting towards a general
similarity of the physics of individual flare events and the overall
radio--X-ray connection. Although the X-ray data do not cover the
start of the X-ray flare, explaining the rather large uncertainty of
the lag determination, the morphological similarities between the
X-ray and radio lightcurves also suggest that the same event is
observed in both wavebands.

Similar events in microquasars show that lags with timescales of
several 100\,s are typical for the coherent behavior of these systems,
such as a lag of 310$\pm$20\,s between the soft X-rays and the IR
\citep{eikenberry:98a} and 800\,s between the X-rays and the radio
\citep{pooley:97a} in GRS~1915+105. The timescale observed in Cyg~X-1
allows us to place an upper limit to the physical separation of the
X-ray and radio emitting regions of the accretion/ejection flow.  We
assume that the emission coincides with the imaged jet and that the
jet is perpendicular to the orbital plane of the HDE~226868/Cyg~X-1
system \citep[although this is not a priori
certain;][]{maccarone:02d}, which has an inclination of $30^\circ$
\citep{gies:82a,dolan:92a}. Taking light travel time effects into
account, for jet speeds of $0.3\,c$, the lower limit implied by
observations of the transient jet ejection discussed by
\citet{fender:06a}, the measured delay implies a separation of $1.1\pm
0.5$\,AU between the location of X-ray and radio emission. If the jet
is relativistic instead, with a speed of $0.99c$, the distance
increases to $5.8\pm2.3$\,AU. Note that similar values for the length
of the jet are obtained by considering that the $\sim$10\,minute
duration of the radio flare is roughly equal to the dynamical
timescale of the jet.  Assuming a distance of 2.5\,kpc, these values
imply to a maximum projected angular separation between the X-ray and
radio emitting region of $10^{-3}$\,mas.

What is the physics of the observed event? In the model of
\citet{fender:04b} for transient radio events, the inner edge of a
thin accretion disk is posited to move rapidly towards the black hole.
The temperature at the inner edge therefore increases, leading to a
softening of the source in the \textsl{RXTE}-PCA as more disk photons
enter the instrument's band pass. This X-ray flare is then followed by
the ejection of an electron bubble, which rapidly expands, producing
the observed radio emission. At least qualitatively, this behavior and
also the time scales deduced above seem to agree with our
observations, although the model was originally invented for the large
scale variability of black hole candidates and not for such short
events as the one discussed here.  Note that while the main flare
dominates the measured time lag and therefore the sizes estimated
above, it is followed by two short spikes, which are both present in
the radio and the X-ray lightcurves, but at different time delays.
These spikes could indicate that more than one blob of material was
ejected at different speeds but cannot be separated once the blobs
have expanded and their radiation peaks in the radio. Such a behavior
could be typical for flares in Cyg~X-1, since the large radio flare of
Cyg~X-1 from 2005 Feb 20 also shows very little substructure
\citep{fender:06a}.

The lack of further detections of radio--X-ray flares in over
1.5\,Msec of simultaneous radio--X-ray data precludes a more detailed
discussion of the properties of the ejected material from comparing
the source behavior during different flares. The observation of the
flare itself, however, stresses the importance of long-term
multi-wavelength campaigns to detect such rare events which are
necessary to further our insight into the physics of the emission from
black holes.

\acknowledgments

We acknowledge the support of NASA contract NAS5-30720, NASA grant
NNG05GK55G, and a travel grant from the Deutscher Akademischer
Austauschdienst. We thank the \textsl{RXTE} schedulers for their
efforts in making the simultaneous radio--X-ray observations possible
and the Aspen Center for Physics for its hospitality during the early
stages of the preparation of this manuscript.

{\it Facilities:} \facility{RXTE (PCA,HEXTE)}, \facility{Ryle ()}

\end{document}